\documentclass[12pt]{iopart}

\usepackage{graphicx,bbm,amssymb,amsopn}

\newcommand{\un}[1]{{\underline{#1}}}

\newcommand{\xbra}[1]{{( #1 \vert}}
\newcommand{\ket}[1]{{\vert #1 \rangle}}

\newcommand{\xbraket}[2]{( #1 \vert #2 \rangle}

\newcommand{\ii}{ {\rm i} }
\newcommand{\dd}{ {\rm d} }
\newcommand{\ZZ}{\mathbb{Z}}

\newcommand{\CC}{\mathbb{C}}

\newcommand{\bb}{ {\hat b} }
\newcommand{\aaa}{ {\hat a} }
\newcommand{\ab}{\un{\aaa}_0}
\newcommand{\ac}{\un{\aaa}_1}

\newcommand{\LL}{{\hat {\cal L}}}

\newcommand{\mm}[1]{{\mathbf{#1}}}

\def\tr{{\,{\rm tr}\,}}

\def\one{\mathbbm{1}}
\def\re{{\,{\rm Re}\,}}

\def\ness{{\rm NESS}}

\begin{document}

\title[Quantization over boson operator spaces]
{Quantization over boson operator spaces}

\author{Toma\v{z} Prosen$^{1,2}$ and Thomas H. Seligman$^{3,4}$}

\address{$^1$ Department of physics, FMF, University of Ljubljana, Ljubljana, Slovenia\\
$^2$ Department of Physics and Astronomy, University of Potsdam, Potsdam, Germany
$^3$ Instituto de Ciencias F\' isicas, Universidad Nacional Aut\'onoma de M\'exico,  
Cuernavaca, Morelos, M\'exico\\
$^4$ Centro Internacional de Ciencias,  Cuernavaca, Morelos, M\'exico\\
}

\date{\today}

\begin{abstract}
The framework of third quantization -- canonical quantization in the Liouville space - 
is developed for open many-body bosonic systems. We show how to diagonalize the quantum Liouvillean 
for an arbitrary quadratic $n-$boson Hamiltonian with arbitrary linear Lindblad couplings to the baths
and, as an example, explicitly work out a general case of a single boson.
\end{abstract}

\maketitle

\section{Introduction}

Liouville spaces of operators have proven very useful in the analysis of many-body problems \cite{fano}, and in particular in optical spectroscopy \cite{general}. In such approaches one equips the Liouville space of quantum operators (representing observables or density matrices) with an inner product such that the Liouville space in fact becomes a Hilbert space.
Therefore, for many body problems, a second quantization over such spaces is tempting and indeed was  presented by one of 
the authors \cite{njp}, and coined as {\it third quantization}, for the finite dimensional case of Fermi operators. 
It was successfully used to explicitly and efficiently solve situations with quadratic (or quasi-free) Hamiltonians and linear coupling to an environment via the Lindblad operators \cite{njp,prl} or via the Redfield model \cite{njp2}. 
In particular, the method already provided exciting new physics, such as a discovery of a quantum phase transition far from equilibrium in XY spin chain \cite{prl}.

The present communication deals with developing a similar framework for bosons. This is of considerable importance for several reasons. 
First, there exists a somewhat dated but vast literature on harmonic oscillators 
and more general quadratic Hamiltonians reported in two books \cite{Malkin, Moshinsky}
essential to the development of quantum optics and many other fields. Second, oscillators and linear couplings 
to the environment are widely used in the theory of decoherence \cite{davidovich}.
Finally, the development of technique of third quantization over infinite-dimensional boson spaces 
seems essential for a future development of a supersymmetric theory of open many-body 
systems involving fermions, bosons, their interaction, and coupling to the environment. 
For quadratic interactions and linear coupling to the environment closed form solutions 
will likely emerge, whereas for non-linearly interacting systems perturbative and non-perturbative methods of many-body physics can be used in conjunction with our approach. 
Though many aspects of the open bosonic problems with linear environment coupling operators have been 
treated before, the elegance and versatility  of the third quantization treatment distinguishes this treatment from others.
 
\section{Preliminaries}
 
Let us consider a Hilbert-Fock space ${\cal H}$ of $n$ {\em bosons}. The representation of ${\cal H}$ 
can be generated by a special element
$\psi_0\in{\cal H}$, called {\em a vacuum pure state}, and a set of $n$ unbounded operators over 
${\cal H}$, $a_1,\ldots,a_n$, and their Hermitian adjoints
$a^\dagger_1,\ldots,a^\dagger_n$, satisfying {\em canonical commutation relations} (CCR)
\begin{equation}
[a_j,a^\dagger_k] = \delta_{j,k}, \quad [a_j,a_k] = [a^\dagger_j,a^\dagger_k] = 0.
\label{eq:CCR}
\end{equation}
Let us define a pair of vector spaces ${\cal K}$ and ${\cal K}'$, such that ${\cal K}$ contains trace class operators, for example density matrices, and
${\cal K}'$ contains unbounded operators which we need as physical observables. 
After choosing a specific space of observables ${\cal K'}$, we define the space ${\cal K}$ as a subspace of 
{\em trace class} operators over ${\cal H}$, such that $\rho \in {\cal K}$ if and only if
$A \rho$ is trace class for any $A \in {\cal K}'$. In this sense ${\cal K}'$ is a dual space to ${\cal K}$.

For instance, we may chose ${\cal K}'$ as a linear space of all ({\em unbounded}) operators 
whose Weyl symbol, i.e the phase space representation of the operator, is an entire function on $\CC^{2n}$.
Then ${\cal K}$ must be limited to operators with finite support in the number operator basis, i.e. to operators which have a finite number of non-vanishing matrix elements in this basis. Such a constraint on density matrices may be too restrictive for certain applications. We shall show later [using eq. (\ref{eq:bases})] that it can be relaxed by restricting ${\cal K}'$.
 
Since the development in the paper will be purely algebraic, we shall adopt Dirac notation and write an element of ${\cal K'}$ as {\em ket} $\ket{\rho}$ and an 
element of ${\cal K}$ as {\em bra} $\xbra{A}$, such that their contraction gives the
expectation value of $A$ for a state $\rho$,
\begin{equation}
\xbraket{A}{\rho} = \tr A\rho.
\label{eq:spr}
\end{equation}
We keep distinct styles of brackets to emphasize the manifest difference between the spaces from
which the ket and the bra have to be chosen.

If $b$ is any of $a_j,a^\dagger_j$, then for each $\rho \in {\cal K}$ and $A \in {\cal K}'$, 
$b \rho, \rho b$ and $A b, b A$ are also elements of ${\cal K}$ and ${\cal K}'$ respectively,
so one can define the left and right multiplication maps $\hat{b}^{\rm L}$ and $\hat{b}^{\rm R}$ over ${\cal K}$
\begin{equation}
\hat{b}^{\rm L} \ket{\rho} = \ket{b \rho},\quad \hat{b}^{\rm R} \ket{\rho} = \ket{\rho b}.
\label{eq:left}
\end{equation}
The action of their adjoint maps on ${\cal K}'$ follows from definition (\ref{eq:spr}) and cyclicity of trace,
\begin{equation}
\xbra{A} \hat{b}^{\rm L}  = \xbra{A b},\quad \xbra{A} \hat{b}^{\rm R} = \xbra{b A}.
\label{eq:right}
\end{equation}
With a slight abuse of notation, we can also write ${(\hat{b}^{\rm L})}^* = \hat{b}^{\rm R}$ and ${(\hat{b}^{\rm R})}^* = \hat{b}^{\rm L}$.

Let us now define the set of $4n$ maps $\hat{a}_{\nu,j},\hat{a}'_{\nu,j},j=1,\ldots,n,\nu=0,1$,
\begin{eqnarray}
\hat{a}_{0,j} &=& \hat{a}^{\rm L}_j ,\,\;\qquad   \hat{a}'_{0,j} =  \hat{a^\dagger}^{\rm L}_j -  \hat{a^\dagger}^{\rm R}_j, \cr
\hat{a}_{1,j} &=& \hat{a^\dagger}^{\rm R}_j ,\qquad   \hat{a}'_{1,j} =  \hat{a}^{\rm R}_j -  \hat{a}^{\rm L}_j.
\label{eq:maps}
\end{eqnarray}
satisfying the unique properties: (i) almost-CCR
\begin{equation}
[\hat{a}_{\nu,j},\hat{a}'_{\mu,k}] = \delta_{\nu,\mu}\delta_{j,k}, \quad [\hat{a}_{\nu,j},\hat{a}_{\mu,k}] =  [\hat{a}'_{\nu,j},\hat{a}'_{\mu,k}]  = 0,
\label{eq:almostCCR}
\end{equation}
(ii) $\hat{a}'_{\nu,j}$ left-annihilate the identity operator
\begin{equation}
\xbra{1}\hat{a}'_{\nu,j} = 0
\label{eq:leftvacuum}
\end{equation}
and (iii) $\hat{a}_{\nu,j}$ right-annihilate the vacuum pure state $\ket{\rho_0} \equiv |\psi_0\!\!> <\!\!\psi_0|$
 \begin{equation}
\hat{a}_{\nu,j}\ket{\rho_0} = 0.
\label{eq:rightvacuum}
\end{equation}
Writing a $2n$ component multi-index $\un{m}=(m_{\nu,j}\in\ZZ_+;\nu\in\{0,1\},j\in\{1\ldots n\})^T$ we can define convenient dual-Fock bases of 
the spaces ${\cal K},{\cal K}'$ as
\begin{equation}
\ket{\un{m}} = \prod_{\nu,j} \frac{(\hat{a}'_{\nu,j})^{m_{\nu,j}}}{\sqrt{m_{\nu,j}!}} \ket{\rho_0},\quad
\xbra{\un{m}} = \xbra{1}\prod_{\nu,j} \frac{(\hat{a}_{\nu,j})^{m_{\nu,j}}}{\sqrt{m_{\nu,j}!}} 
\label{eq:bases}
\end{equation}
whose bi-orthonormality $\xbraket{\un{m}'}{\un{m}} = \delta_{\un{m}',\un{m}}$ is simply guaranteed by almost-CCR (\ref{eq:almostCCR}).
Here and below, $\un{x}=(x_1,x_2,\ldots)^T$ designates a vector (column) of any, scalar-, operator- or map-valued symbols, 
$\un{x}\cdot\un{y}=x_1 y_1 + x_2 y_2 + \ldots$ designates a dot product, and bold upright letters $\mm{A}$ shall be used for complex valued matrices. 

The explicit construction of the bases (\ref{eq:bases}) allows us to enlarge the space ${\cal K}$ and restrict the space ${\cal K}'$ as compared to the above example.
Namley, we now identify the space ${\cal K}$ with the $l^2$ Hilbert space of vectors of coefficients $\{ \sigma_{\un{m}} \}$, 
${\cal K} \ni \ket{\sigma} = \sum_{\un{m}} \sigma_{\un{m}} \ket{\un{m}}$, and the space ${\cal K}'$ with the $l^2$ Hilbert space of vectors of coefficients
$\{ S_{\un{m}} \}$, ${\cal K}' \ni \xbra{S} = \sum_{\un{m}} S_{\un{m}} \xbra{\un{m}}$. Then, clearly by Cauchy-Schwarts inequality, 
$|\tr S\sigma| = |\sum_{\un{m}} S_{\un{m}}\sigma_{\un{m}}| < \infty$ and hence ${\cal K}$ and ${\cal K}'$ are dual in the required sense.
 
\section{Explicit solution of the Lindblad equation for quadratic bosonic systems}

Our goal is to present an exact solution of the master equation of an open $n$-particle system, say of the Linblad form 
\begin{equation}
\frac{\dd\rho }{\dd t} = \LL\rho :=
-\ii [H,\rho] + \sum_{\mu} \left(2 L_\mu \rho L_\mu^\dagger - \{L_\mu^\dagger L_\mu,\rho\} \right)
\label{eq:lind}
\end{equation}
where $H$ is a Hermitian operator (Hamiltonian), $\{x,y\}:=xy+yx$, and $L_\mu$ are arbitrary (Lindblad) operators representing couplings to different
baths. We are going to describe a general method of explicit solution of (\ref{eq:lind}) for an arbitrary {\em quadratic} system of $n$ bosons
with {\em linear} bath operators
\begin{eqnarray}
H &=& \un{a}^\dagger \cdot \mm{H} \un{a} + \un{a} \cdot \mm{K}\un{a} + \un{a}^\dagger\cdot \mm{\bar K}\un{a}^\dagger \label{eq:hamil}  \\
L_\mu &=& \un{l}_\mu \cdot \un{a} + \un{k}_\mu \cdot \un{a}^\dagger.  \label{eq:lindb}
\end{eqnarray}
Note that the hermiticity of the Hamiltonian and CCR (\ref{eq:CCR}) implies that the matrix $\mm{H}$ is Hermitean $\mm{H}^\dagger = \bar{\mm{H}}^T = \mm{H}$ and the matrix $\mm{K}$ is symmetric 
$\mm{K}=\mm{K}^T$. For bound oscillator systems, one can always choose $\mm{K}=0$, however, in order to be able to describe freely moving or inverted oscillator modes, we shall keep the general form.

Following the rules of the algebra (\ref{eq:left},\ref{eq:right},\ref{eq:maps}) we straightforwardly express the Liouvillean (\ref{eq:lind}) in terms of the canonical maps
\begin{eqnarray}
\LL &=& -\ii \hat{H}^{\rm L} + \ii \hat{H}^{\rm R} + \sum_\mu 2\hat{L_\mu^{\phantom{\dagger}}}^{\rm L} \hat{L^\dagger_\mu}^{\rm R} - \hat{L^\dagger_\mu}^{\rm L}\hat{L_\mu^{\phantom{\dagger}}}^{\rm L} 
- \hat{L_\mu^{\phantom{\dagger}}}^{\rm R}\hat{L_\mu^\dagger}^{\rm R} \\
&=&-\ii \un{\aaa}'_0\cdot \mm{H}\un{\aaa}_0 + \ii \un{\aaa}'_1\cdot \mm{\bar H}\un{\aaa}_1 + \ii \un{\aaa}'_1 \cdot\mm{K} (2\un{\aaa}_0+\un{\aaa}'_1) - \ii \un{\aaa}'_0 \cdot 
\mm{\bar K}(2\un{\aaa}_1+\un{\aaa}'_0) \nonumber \\
&+&  \ab'\cdot(\mm{N}\!-\!\mm{\bar M})\ab +  \ac'\cdot(\mm{\bar N}\!-\!\mm{M})\ac 
+ \ab'\cdot(\mm{\bar L}^T\!-\!\mm{\bar L})\ac +\ac'\cdot(\mm{L}^T\!-\!\mm{L})\ab \nonumber \\
&-& \ab'\cdot\mm{\bar L}\ab' - \ac'\cdot\mm{L}\ac' + 2\ab'\cdot\mm{N}\ac'
\label{eq:liouv1}
\end{eqnarray}
where we define $n \times n$ matrices
\begin{equation}
\mm{M} := \sum_\mu \un{l}_\mu \otimes \un{\bar l}_\mu = \mm{M}^\dagger, \; \mm{N} := \sum_\mu \un{k}_\mu \otimes \un{\bar k}_\mu = \mm{N}^\dagger,\;
\mm{L}:= \sum_\mu \un{l}_\mu \otimes \un{\bar k}_\mu.
\end{equation}
The fact that the primed maps $\un{\aaa}'$ always appear on the left side in each term manifestly expresses the {\em trace preservation} of the flow (\ref{eq:lind}), i.e. $\xbra{1}\LL = 0$.

Note that adding inhomogeneous terms (linear forces) to the Hamiltonian $H \to H + \un{f}\cdot\un{a} + \un{\bar f}\cdot \un{a}^\dagger$ and the
Lindblad operators $L_\mu \to L_\mu + \lambda_\mu \hat{\one}$, where $\un{f}$ is a c-vector and $\lambda_\mu$ are c-numbers,
simply results in an extra {\em linear} term in the Liouvillean $\LL \to \LL + \un{g}\cdot \un{\aaa}'$, which must be removed by a {\em canonical transformation} affecting the
annihilation maps $\hat{a}_{\nu,j}$ only, namely $\un{\aaa} \to  \un{\aaa}  + \un{s}\one,\un{\aaa}'\to \un{\aaa}'$ and thus preserving the important relations (\ref{eq:almostCCR},\ref{eq:leftvacuum}), while 
the right vacuum state  $\ket{\rho_0}$ (which is unimportant for the following discussion) is trivially shifted.

Writing $4n$ vector of canonical maps as $\un{\bb} = (\un{\aaa},\un{\aaa}')^T = (\ab,\ac,\ab',\ac')^T$ the 
Liouvillean (\ref{eq:liouv1}) can be compactly rewritten in terms of a symmetric form
\begin{equation}
\LL = \un{\bb}\cdot \mm{S} \un{\bb} - S_0 \hat{\one}
\label{eq:liouv2}
\end{equation}
where $\mm{S}$ is a complex symmetric $4n \times 4n$ matrix which can be conveniently written in terms of two $2n\times 2n$ matrices
\begin{equation}
\mm{S} = \pmatrix{ \mm{0} & \mm{-\mm{X}} \cr
-\mm{X}^T & \mm{Y}},
\end{equation}
namely
\begin{equation}
\mm{X} := \frac{1}{2}\pmatrix{ \ii \mm{\bar H}-\mm{\bar N}+\mm{ M} & -2\ii\mm{K}-\mm{L}+\mm{L}^T \cr
2\ii\mm{\bar K}-\mm{\bar L}+\mm{\bar L}^T & -\ii \mm{H} - \mm{N} + \mm{\bar M}}
\label{eq:X}
\end{equation}
and
\begin{equation}
\mm{Y} :=\frac{1}{2}\pmatrix{-2\ii\mm{\bar K} - \mm{\bar L}-\mm{\bar L}^T & 2\mm{N} \cr 2\mm{N}^T & 2\ii \mm{K} - \mm{L} - \mm{L}^T} = \mm{Y}^T.
\label{eq:Y}
\end{equation}
The scalar $S_0$ (\ref{eq:liouv2}) stemming from reordering of maps equals $S_0 = \tr \mm{X} = \tr\mm{M} - \tr\mm{N}$.

For the rest of our discussion we shall assume that the matrix $\mm{X}$ is diagonalizable \cite{nondiag}, i.e. it is similar to a diagonal matrix 
\begin{equation}
\mm{X} = \mm{P} \mm{\Delta} \mm{P}^{-1},\quad \mm{\Delta} = {\rm diag}\{\beta_{1},\ldots,\beta_{2n}\}
\label{eq:diag}
\end{equation}
where the $2n$ complex eigenvalues $\beta_j$ shall be named as {\em rapidities} in analogy to the fermionic case \cite{njp}. Let 
$\mm{J} = \ii\sigma_{\rm y} \otimes \one_{2n}$ denote the symplectic unit, satisfying $\mm{J}^2 = -\one_{4n}$. 
It is straightforward to check that the matrix $\mm{J}\mm{S}$ belongs to the {\em symplectic algebra}, namely $\mm{J}(\mm{J}\mm{S}) = -(\mm{J}\mm{S})^T \mm{J}$, and
can be diagonalized as
\begin{equation}
\mm{J}\mm{S} = \mm{V}^{-1} [(-\mm{\Delta}) \oplus \mm{\Delta}] \mm{V},
\quad 
\mm{V} = [\mm{P}^T \oplus \mm{P}^{-1}] \pmatrix{ \one_{2n} & -\mm{Z} \cr \mm{0} & \one_{2n}} \label{eq:decomp}
\end{equation}
with the eigenvector matrix $\mm{V}$ belonging to the complex symplectic group
\begin{equation}
\mm{V}^T \mm{J} \mm{V} = \mm{J}.
\label{eq:symp}
\end{equation}
The $2n\times 2n$ complex symmetric matrix $\mm{Z}=\mm{Z}^T$ is a solution of the {\em continuous Lyapunov equation} \cite{lyap}
\begin{equation}
\mm{X}^T \mm{Z} + \mm{Z}\mm{X} = \mm{Y}.
\label{eq:lyap}
\end{equation}
It is known that the solution of (\ref{eq:lyap}) exists and is unique if no pairs of eigenvalues, rapidities $\beta_j,\beta_{j'}$ exist, such that $\beta_j + \beta_{j'}=0$.
We note that $\mm{X}$,  and also $\mm{Y}$, are (unitarily) similar to real matrices, namely if $\mm{U} := \frac{1}{\sqrt{2}}(\one_2 + \ii \sigma_{\rm x})\otimes \one_{2n}$ then
$\mm{U}\mm{X}\mm{U}^{-1}$, and $\mm{U}\mm{Y}\mm{U}^{-1}$, are {\em real} matrices.\footnote{In fact, if we were working in terms of Hermitian ``coordinate'' and ``momentum" operators,
$a_j + a_j^\dagger, \ii (a_j - a^\dagger_j)$, from the outset, then a very similar formalism could have been developed with matrices $\mm{X}$ and $\mm{Y}$ being automatically real.}
Thus the rapidities should come in complex conjugate pairs $\beta_j,\bar{\beta}_j$. 
Existence and uniqueness of the solution of (\ref{eq:lyap}) and further, the decomposition (\ref{eq:decomp}), is thus guaranteed if all the rapidities lie {\em away from the imaginary axis}, $\re \beta_j \neq 0$.
Diagonalizing $\mm{X}$ (\ref{eq:diag}) is then also the essential part of efficiently computing the solution $\mm{Z}$ of (\ref{eq:lyap}) \cite{sylv}.

Defining the $4n$ {\em normal master-mode maps} (NMM) as $(\un{\hat\zeta},\un{\hat\zeta}')^T := \mm{V} \un{\hat b}$, or
\begin{equation}
\un{\hat\zeta} = \mm{P}^{T}(\un{\aaa} - \mm{Z}\un{\aaa}'),\quad
\un{\hat\zeta}' = \mm{P}^{-1}\un{\aaa}'
\label{eq:NMM}
\end{equation}
which, due to symplecticity (\ref{eq:symp}), satisfy almost-CCR
\begin{equation}
[\hat{\zeta}_r,\hat{\zeta}'_s] = \delta_{r,s},\quad [\hat{\zeta}_r,\hat{\zeta}_s] =[\hat{\zeta}'_r,\hat{\zeta}'_s] =0,
\end{equation}
brings the Liouvillean to the normal-form
$\LL  = \mm{J}\un{\hat b}\cdot \mm{J}\mm{S}\un{\hat b} - S_0 \hat{\one}  = \mm{J}\un{\hat\zeta}\cdot  [(-\mm{\Delta}) \oplus \mm{\Delta}] \un{\hat\zeta} - S_0 \hat{\one} $ 
and finally
\begin{equation}
\LL = -2\sum_{r=1}^{2n} \beta_r \hat{\zeta}'_r \hat{\zeta}_r.
\end{equation}

In contrast to the (finite dimensional) fermionic case \cite{njp}, existence of a stable fixed point of the Liouvillean dynamics is here not guaranteed. Since our (central) system is infinitely dimensional, it can 
absorb excitations from the environment and amplify indefinitely. Indeed, this is signaled by at least one of the rapidities being to the left of the imaginary axis, $ \exists j, \re \beta_j < 0$.

However, if we assume that all the rapidities lie to the right of the imaginary axis, $\forall j, \re \beta_j > 0$, then we can make the following statements:
\begin{enumerate}
\item A {\em unique} non-equlibrium steady state (NESS) exists $\ket{\ness} \in{\cal K}$ 
namely the ``right vacuum state" of the Liouvillean
\begin{equation}
\LL\ket{\ness} = 0.
\end{equation}
All the physical properties of NESS are essentially determined by the NMM annihilation relations 
\begin{equation}
\xbra{1}\hat{\zeta}'_r=0,\quad \hat{\zeta}_r \ket{\ness}=0.
\label{eq:anih}
\end{equation}
\item
The complete (point) spectrum of the decay modes of the Liouvillean is given in terms of a $2n$ component multi-index of super-quantum numbers $\un{m}\in \ZZ_+^{2n}$
\begin{equation}
\lambda_{\un{m}} = -2\sum_r m_r \beta_r,\quad
\LL \ket{\Theta_\un{m}} = \lambda_{\un{m}} \ket{\Theta_\un{m}},\quad 
\xbra{\Theta_\un{m}} \LL = \lambda_{\un{m}}\xbra{\Theta_\un{m}}
\end{equation}
where
\begin{equation}
\ket{\Theta_\un{m}} = \prod_{r} \frac{(\hat{\zeta}'_r)^{m_r}}{\sqrt{m_r!}} \ket{\ness},
\quad \xbra{\Theta_\un{m}} = \xbra{1}\prod_{r} \frac{(\hat{\zeta}_r)^{m_r}}{\sqrt{m_r!}}.
\end{equation}
\item
The 2-point correlator of NESS is given by the solution of the Lyapunov equation (\ref{eq:lyap}).
If $\un{b} = (\un{a},\un{a}^\dagger)^T$ designates a $2n$ column of canonical operators  then
\begin{equation}
\tr :\! b_r b_s\! : \rho_{\ness} =  \xbraket{:\! b_r b_s\! :}{\ness} = \xbra{1} \aaa_r \aaa_s\ket{\ness} = Z_{r,s}.
\label{eq:observable}
\end{equation}
\end{enumerate}
{\em Proof.} Using purely algebraic manipulations (i) and (ii) can be proved in analogy to the fermionic case \cite{njp} and are in fact equivalent to the standard construction of the Fock ground state and quasi-particle excitations 
with a restriction of non-normality.
(iii) The first two equality signs are just the definitions (\ref{eq:spr},\ref{eq:maps}) whereas the last equality follows after expressing $\aaa_r$ in terms of NMM maps (\ref{eq:NMM}), and using (\ref{eq:anih}).

We note that NESS is a essentially a Gaussian state and Wick theorem can be used to express any higher-order correlator in terms of 2-point contractions (\ref{eq:observable}).

\section{Example: a general Linbldad equation for a single quantum oscillator}

\label{sect:example}

Take the simplest case, namely a single quantum harmonic oscillator $n=1$. We choose $\mm{H}\equiv\omega, \mm{K}\equiv0,
\mm{M} \equiv u = \sum_\mu |l_\mu|^2 > 0$, $\mm{N} \equiv v = \sum_\mu |k_\mu|^2 > 0$, $\mm{L} \equiv w = \sum_\mu l_\mu \bar{k}_\mu \in \CC$. By construction the reservoir parameters should satisfy the positivity condition $ u v > |w|^2$.
The matrices (\ref{eq:X},\ref{eq:Y}) now read
\begin{equation}
\mm{X}= \frac{1}{2}\pmatrix{ \ii \omega + u - v & 0 \cr
0 & -\ii \omega + u - v},\quad
\mm{Y} = \pmatrix{-\bar{w} & v \cr v & -w}.
\end{equation}
The rapidity spectrum $\beta_{\pm} = \frac{1}{2}(u - v \pm \ii\omega)$ indicates that the problem has a stable fixed point if and only if $u > v$.
Then, NESS is approached at an exponential rate $\sim \exp[-(u-v)t]$ and the complete corelator (\ref{eq:observable}), as given by the straightforward solution of the Lyapunov equation (\ref{eq:lyap}) (now a linear system of $3$ equations), read
\begin{eqnarray}
\tr a^2 \rho_{\ness} &=& Z_{11} = \frac{\bar{w}}{u-v+\ii \omega} =  \overline{ \tr (a^\dagger)^2 \rho_{\ness}}, \nonumber \\
\tr a^\dagger a \rho_{\ness} &=& Z_{12} =  \frac{v}{u-v}.
\end{eqnarray}
It is perhaps worth noting that this is {\em not} a Gibbs thermal state for any combination of Lindblad parameters, the fact which can be viewed as a manifestation of non-ergodicity of the harmonic oscillator. Furhtermore, since NESS is a generalized Gaussian state, expectation values of all higher moments of the field are calculated trivially.

\section{Discussion}

In the present communication we outlined a simple and mathematically consistent framework for diagonalizing quantum Liouvilleans for arbitrary bi-linear systems of bosons.
Our method in some sense resembles the technique known as {\em non-equilibrium thermo-field dynamics} \cite{umezawa,arimitsu}, which has been recently applied to a similar problem of
diagonalizing the Lindblad equation \cite{ban}. However, there is an important difference, namely in thermo-field dynamics, one treats a density matrix as an element of a tensor product Hilbert space, whereas the
operator (observable) of choice is fixed in a given calculation. In our approach, we consider (dual) spaces of both simultaneously, so the canonical structure of our formalism is simpler and more transparent. This, in turn, is reflected in the ease of getting simple general results  as outlined in the example in section \ref{sect:example}).

Notice that our formalism is flexible with respect to taking different realizations of the dual spaces ${\cal K}$ and ${\cal K}'$, namely for some physics
applications it may be advantageous to take larger space of observables ${\cal K}'$ at the expense of a smaller dual space ${\cal K}$, or vice versa.
In analogy to the fermionic case \cite{njp2} this formalism can be extended to explicitly time dependent problems. For example, for periodic time dependent problems \cite{diploma}, the Floquet representation of the correlation matrix leads to a {\em discrete Lyapunov equation} rather than a {\em continuous} one that appears here.
 
The approach presented here differs essentially from the framework proposed earlier for fermionic systems \cite{njp}, where the density operators and the observables belong to the same Hilbert space. For bosonic systems, due to infinite-dimensionality, this is not possible and the symmetry between density operator space and observable operator space is broken. However, it may be desirable for certain purposes to have formally similar construction for bosonic and fermionic systems.
Thus we point out an equivalent version of fermionic third quantization, which follows exactly the steps of the present communication, but starting instead from a set of fermionic operators $c_j, c^\dagger_j$, obeying
{\em canonical anti-commutation relations} (CAR),
\begin{equation}
\{c_j,c^\dagger_k\}= \delta_{j,k}, \quad \{c_j,c_k\} = \{c^\dagger_j,c^\dagger_k\} = 0,
\label{eq:CAR}
\end{equation}
introducing the dual spaces of density operators and observables, stating (\ref{eq:spr},\ref{eq:left},\ref{eq:right}) and defining the canonical adjoint fermionic maps
\begin{eqnarray}
\hat{c}_{0,j} &=& \hat{c}^{\rm L}_j ,\;\quad\qquad   \hat{c}'_{0,j} =  \hat{c^\dagger}^{\rm L}_j -  \hat{c^\dagger}^{\rm R}_j \hat{\cal P}, \cr
\hat{c}_{1,j} &=& \hat{c^\dagger}^{\rm R}_j \hat{\cal P} ,\qquad   \hat{c}'_{1,j} =  \hat{c}^{\rm R}_j \hat{\cal P} -  \hat{c}^{\rm L}_j,
\label{eq:fmaps}
\end{eqnarray}
satisfying almost-CAR
\begin{equation}
\{\hat{c}_{\nu,j},\hat{c}'_{\mu,k}\} = \delta_{\nu,\mu}\delta_{j,k}, \quad \{\hat{c}_{\nu,j},\hat{c}_{\mu,k}\} =  \{\hat{c}'_{\nu,j},\hat{c}'_{\mu,k}\} = 0,
\label{eq:almostCAR}
\end{equation}
and the properties (\ref{eq:leftvacuum},\ref{eq:rightvacuum}). The parity superoperator $\hat{\cal P}$ is uniquely defined by the dual vacuum states, $\xbra{1}\hat{\cal P} = \xbra{1}, \hat{\cal P}\ket{\rho_0} = \ket{\rho_0}$, and anticommutes with all elements of the adjoint-algebra $\{\hat{\cal P},\un{\hat c}\} = \{\hat{\cal P},\un{\hat c}'\} = 0$.
The difference to the more symmetric approach \cite{njp} is that now the canonical conjugate adjoint maps are {\em not} the hermitian adjoint maps
$\hat{c}'_{\nu,j} \neq \hat{c}^\dagger_{\nu,j}$, which is however of no consequence as we are anyway dealing with problems in which
{\em non-normal} operators enter in an essential way.
 
\section*{Acknowledgments}

We acknowledge discussions with F. Leyvraz and J. Eisert.
This work was supported by the Programme P1-0044, and the Grant J1-2208, of Slovenian Research Agency, 
and by CONACyT, Mexico, project 57334 as well as the University of Mexico, PAPIIT project IN114310.


\begin{thebibliography}{10}
\bibitem{nondiag}  The more general case allowing for a possible non-generic non-diagonalizability of $\mm{X}$ shall be treated in future work.
\end{thebibliography}
\end{document}